# Consider Uncertain Parameters based on Sensitivity Matrix

Taishan Lou

School of Electric and Information Engineering, Zhengzhou University of Light Industry, Zhengzhou, 45002, China, tayzan@sina.com

**Abstract**

Uncertain parameters of state-space models have always been a considerable problem. Consider Kalman filter (CKF) and desensitized Kalman filter (DKF) are two methods to solve this problem. Based on the sensitivity matrix respected to the uncertain parameter vector, a special DKF with an analytical gain is given and a new form of the CKF is derived. The mathematical equivalence between the special DKF and the CKF is demonstrated when the sensitivity-weighting matrix is set to the covariance of the uncertain parameter and the problem how to select and obtain the sensitivity-weighting matrix in the DKF is solved.

## 1. Introduction

An essential assumption on the Kalman filter theory is that the dynamic models can be accurately modeled without any colored noises or uncertain parameters. However, the performance of the Kalman filter can deteriorate significantly due to uncertain parameters in the dynamic models (Schmidt, 1966; Sayed & Chandrasekaran, 2000). The consider Kalman filter (CKF), which is also known as Schmidt-Kalman filter (Schmidt, 1966), is proposed to consider the uncertain parameters by updating the state estimates and covariance using the a priori parameter covariance without estimating them directly (Tapley, et al., 2004; Woodbury & Junkins, 2010). Meanwhile, the desensitized Kalman filter (DKF) based on the desensitized optimal control technique is proposed to decrease the sensitivity to deviations in the uncertain parameters (Karlgaard & Shen, 2011). In case of DKF, the sensitivity of the state estimate error respected to the uncertain parameter is defined to penalty the cost function of the traditional Kalman filter (KF) (Crassidis & Junkins, 2012), and the relevant gain matrix is obtained by minimizing the new cost function with the sensitivity of the state estimate error.

In this work, the sensitivity matrix is introduced into the CKF and obtained a novel form CKF. Based on some results between the cross-covariance matrix in the CKF and the sensitivity matrix in the DKF, the equivalence between the CKF and a special DKF is demonstrated and solving the selection problem of the sensitivity weighting matrix in the DKF.

## 2. Dynamic models with uncertain parameters

Consider a discrete-time linear dynamic system with uncertain parameters:

$$\mathbf{x}_k = \mathbf{\Phi}_{k/k-1}\mathbf{x}_{k-1} + \mathbf{\Psi}_{k/k-1}\mathbf{p} + \mathbf{G}_{k-1}\mathbf{w}_{k-1} \quad (1)$$

$$\mathbf{z}_k = \mathbf{H}_k\mathbf{x}_k + \mathbf{N}_k\mathbf{p} + \mathbf{v}_k \quad (2)$$

where $\mathbf{x}_k$ is the $n \times 1$ state vector, $\mathbf{z}_k$ the $m \times 1$ measurement vector, $\mathbf{p}$ the $\ell \times 1$ uncertain constant parameter vector. $\mathbf{\Phi}_{k/k-1}$ the state transition matrix, $\mathbf{H}_k$ is the measurement matrix, $\mathbf{\Psi}_{k/k-1}$ and $\mathbf{N}_k$ are the deterministic parameter transition matrices, and $\mathbf{G}_{k-1}$ is the process matrix. $\mathbf{w}_k$ and $\mathbf{v}_k$ are independent zero-mean Gaussian noise processes, and satisfy

$$E[\mathbf{w}_k\mathbf{w}_j^T] = \mathbf{Q}_k\delta_{kj},\ E[\mathbf{v}_k\mathbf{v}_j^T] = \mathbf{R}_k\delta_{kj},\ E[\mathbf{w}_k\mathbf{v}_j^T] = \mathbf{0} \quad (3)$$

where "$E$" is the expectation symbol, $\delta$ the Kronecker delta function, and $\delta_{kj}=1$ if $k=j$ and $\delta_{kj}=0$ if $k \neq j$.

In this work, the uncertain parameter $\mathbf{p}$, which is also called bias (Schmidt, 1966; Zanetti & Bishop, 2012), is model as a random constant vector with the a priori statistical properties

$$E[\mathbf{p}] = \hat{\mathbf{p}},\ E[(\hat{\mathbf{p}}-\mathbf{p})(\hat{\mathbf{p}}-\mathbf{p})^T] = \mathbf{P}_{pp}, \quad (4)$$

and $\mathbf{p}$ is independent with the process and measurement noises followed by

$$E[\mathbf{v}_k(\hat{\mathbf{p}}-\mathbf{p})^T] = \mathbf{0}, \; E[\mathbf{w}_k(\hat{\mathbf{p}}-\mathbf{p})^T] = \mathbf{0} \quad (5)$$

where $\hat{\mathbf{p}}$ is reference value of the uncertain parameter $\mathbf{p}$.

## 3. Consider Kalman filter

The CKF, which is proposed by Schmidt to account for the uncertain parameters, incorporates the a priori statistics of the uncertain parameters into the system formulations, and discarding the estimation of the uncertain parameters (Schmidt, 1966; Woodbury & Junkins, 2010). In the CKF algorithm, the rows of the augmenting gain matrix corresponding to the uncertain parameters are set to zeros, because the uncertain parameters are not estimated (Jazwinski, 1970). The CKF algorithm of the discrete-time linear dynamic system as in the reference is described briefly here (Woodbury & Junkins, 2010).

Time update:

$$\hat{\mathbf{x}}_k^- = \mathbf{\Phi}_{k|k-1}\hat{\mathbf{x}}_{k-1}^+ + \mathbf{\Psi}_{k/k-1}\hat{\mathbf{p}} \quad (6)$$

$$\mathbf{P}_k^- = \mathbf{\Phi}_{k/k-1}\mathbf{P}_{k-1}^+\mathbf{\Phi}_{k/k-1}^T + \mathbf{\Phi}_{k/k-1}\mathbf{C}_{k-1}^+\mathbf{\Psi}_{k/k-1}^T + \\ \mathbf{\Psi}_{k/k-1}\mathbf{C}_{k-1}^{+T}\mathbf{\Phi}_{k/k-1}^T + \mathbf{\Psi}_{k/k-1}\mathbf{P}_{cc}\mathbf{\Psi}_{k/k-1}^T + \mathbf{G}_{k-1}\mathbf{Q}_{k-1}\mathbf{G}_{k-1}^T \quad (7)$$

$$\mathbf{C}_k^- = \mathbf{\Phi}_{k/k-1}\mathbf{C}_{k-1}^+ + \mathbf{\Psi}_{k/k-1}\mathbf{P}_{pp} \quad (8)$$

Measurement update:

$$\mathbf{K}_k = (\mathbf{P}_k^-\mathbf{H}_k^T + \mathbf{C}_{k-1}^+\mathbf{N}_k^T)\mathbf{\Omega}_k^{-1} \quad (9)$$

$$\hat{\mathbf{x}}_k^+ = \hat{\mathbf{x}}_k^- + \mathbf{K}_k(\mathbf{z}_k - \mathbf{H}_k\hat{\mathbf{x}}_k^- - \mathbf{N}_k\hat{\mathbf{p}}) \quad (10)$$

$$\mathbf{P}_k^+ = (\mathbf{I} - \mathbf{K}_k\mathbf{H}_k)\mathbf{P}_k^- - \mathbf{K}_k\mathbf{N}_k\mathbf{C}_k^{-T} \quad (11)$$

$$\mathbf{C}_k^+ = (\mathbf{I} - \mathbf{K}_k\mathbf{H}_k)\mathbf{C}_k^- - \mathbf{K}_k\mathbf{N}_k\mathbf{P}_{pp} \quad (12)$$

where $\hat{\mathbf{x}}_k^-$ is the state estimate at $t_k$ before a measurement update and $\hat{\mathbf{x}}_{k-1}^+$ is the state estimate after the measurement update at $t_{k-1}$. $\mathbf{\Omega}_k$ in Eq. (9) is

$$\mathbf{\Omega}_k = \mathbf{H}_k\mathbf{P}_k^-\mathbf{H}_k^T + \mathbf{N}_k\mathbf{C}_k^{-T}\mathbf{H}_k^T + \mathbf{H}_k\mathbf{C}_k^-\mathbf{N}_k^T + \mathbf{N}_k\mathbf{P}_{pp}\mathbf{N}_k^T + \mathbf{R}_k \quad (13)$$

$\mathbf{C}_k^-$ is the a priori cross-covariance matrix defined as

$$\mathbf{C}_k^- = E[(\hat{\mathbf{x}}_k^- - \mathbf{x}_k)(\hat{\mathbf{p}} - \mathbf{p})^T] \quad (14)$$

and $\mathbf{C}_k^+$ is the a posteriori cross-covariance matrix defined as

$$\mathbf{C}_k^+ = E[(\hat{\mathbf{x}}_k^+ - \mathbf{x}_k)(\hat{\mathbf{p}} - \mathbf{p})^T] \quad (15)$$

The superscript $^-$ denotes a priori and $^+$ denotes a posteriori.

## 4. Desensitized Kalman filter

The DKF, in which the gain is obtained to solve a linear matrix equation and the sensitivity-weighting matrix must be preset with a priori knowledge, is presented by Karlgaard and Shen (Karlgaard & Shen, 2011). The state error sensitivity respected to single uncertain parameter is defined to penalty the cost function, and the gain matrix is obtained by solving a linear matrix equation with a sensitivity-weighting matrix. Here, the state error sensitivity respected to the uncertain parameter vector is introduced to obtain an analytical gain matrix, which is the same as the KF in the well-known form (Crassidis & Junkins, 2012).

As in the KF, an unbiased state estimate at time $t_k$ can be obtained from the unbiased state $\hat{\mathbf{x}}_{k-1}^+$ by the following propagation equation

$$\hat{\mathbf{x}}_k^- = \mathbf{\Phi}_{k/k-1}\hat{\mathbf{x}}_{k-1}^+ + \mathbf{\Psi}_{k/k-1}\hat{\mathbf{p}} \quad (16)$$

Then, the KF provides an optimal blending of the $\hat{\mathbf{x}}_k^-$ and $\mathbf{z}_k$ to obtain the a posteriori state estimate via

$$\hat{\mathbf{x}}_k^+ = \hat{\mathbf{x}}_k^- + \mathbf{K}_k(\mathbf{z}_k - \mathbf{H}_k\hat{\mathbf{x}}_k^- - \mathbf{N}_k\hat{\mathbf{p}}) \quad (17)$$

The a priori estimation error is defined as $\mathbf{e}_k^- = \hat{\mathbf{x}}_k^- - \mathbf{x}_k$ and the a posteriori estimation error is defined as $\mathbf{e}_k^+ = \hat{\mathbf{x}}_k^+ - \mathbf{x}_k$. Then, the state error sensitivities and propagation equations of the parameter vector $\mathbf{p}$ are given by (Tapley, et al., 2004)

$$\mathbf{S}_k^- = \frac{\partial \mathbf{e}_k^-}{\partial \mathbf{p}} = \frac{\partial \hat{\mathbf{x}}_k^-}{\partial \mathbf{p}} = \mathbf{\Phi}_{k/k-1}\mathbf{S}_{k-1}^+ + \mathbf{\Psi}_{k/k-1} \quad (18)$$

$$\mathbf{S}_k^+ = \frac{\partial \mathbf{e}_k^+}{\partial \mathbf{p}} = \frac{\partial \hat{\mathbf{x}}_k^+}{\partial \mathbf{p}} = \mathbf{S}_k^- - \mathbf{K}_k\mathbf{\gamma}_k \quad (19)$$

where

$$\mathbf{\gamma}_k = \mathbf{H}_k\mathbf{S}_k^- + \mathbf{N}_k. \quad (20)$$

Note that the sensitivity of the true state is $\partial \mathbf{x}_k / \partial \mathbf{p} = 0$ in Eqs. (18) and (19), and it is assumed that $\partial \mathbf{K}_k / \partial \mathbf{p} = 0$ in Eq. (19)(Karlgaard & Shen, 2011). The sensitivity matrices $\mathbf{S}_k^-$ and $\mathbf{S}_k^+$ describe how the state estimate errors $\mathbf{e}_k^-$ and $\mathbf{e}_k^+$ vary with the uncertain parameter vector $\mathbf{p}$.

Based on the definitions of the sensitivity matrices in Eqs. (18) and (19), a new cost function based on the trace of the weighted norm of the a posterior sensitivity matrix is given by

$$J = Tr(\mathbf{P}_k^+) + Tr(\mathbf{S}_k^+ \mathbf{W} \mathbf{S}_k^{+T}) \qquad (21)$$

where "Tr" denotes the trace of the matrix, $W$ is a $\ell \times \ell$ symmetric positive semi-definite weighting matrix for the uncertain parameters, and $\mathbf{P}_k^+$ is the a posteriori error covariance matrix

$$\mathbf{P}_k^+ = E[\mathbf{e}_k^+ \mathbf{e}_k^{+T}] = (\mathbf{I} - \mathbf{K}_k \mathbf{H}_k) \mathbf{P}_k^- (\mathbf{I} - \mathbf{K}_k \mathbf{H}_k)^T + \mathbf{K}_k \mathbf{R}_k \mathbf{K}_k^T \qquad (22)$$

Note that $\mathbf{K}_k$ is the undetermined gain matrix, and $\mathbf{P}_k^-$ is the a priori error covariance matrix

$$\mathbf{P}_k^- = \mathbf{\Phi}_{k/k-1} \mathbf{P}_{k-1}^+ \mathbf{\Phi}_{k/k-1}^T + \mathbf{G}_{k-1} \mathbf{Q}_{k-1} \mathbf{G}_{k-1}^T \qquad (23)$$

Substituting Eqs. (19) and (22) into Eq. (21) and taking the derivative with respect to the gain $\mathbf{K}_k$ by using the trace derivative properties, yields

$$\frac{\partial J}{\partial \mathbf{K}_k} = 2\mathbf{K}_k(\mathbf{H}_k \mathbf{P}_k^- \mathbf{H}_k^T + \mathbf{R}_k) - 2\mathbf{P}_k^- \overline{\mathbf{H}}_k^T - 2\mathbf{S}_k^- \mathbf{W} \boldsymbol{\gamma}_k^T + 2\mathbf{K}_k \boldsymbol{\gamma}_k \mathbf{W} \boldsymbol{\gamma}_k^T \qquad (24)$$

Setting $\partial J / \partial \mathbf{K}_k = 0$, the analytical solution of the gain $\mathbf{K}_k$ is

$$\mathbf{K}_k = (\mathbf{P}_k^- \mathbf{H}_k^T + \mathbf{S}_k^- \mathbf{W} \boldsymbol{\gamma}_k^T)(\mathbf{H}_k \mathbf{P}_k^- \mathbf{H}_k^T + \boldsymbol{\gamma}_k \mathbf{W} \boldsymbol{\gamma}_k^T + \mathbf{R}_k)^{-1} \qquad (25)$$

Substituting Eq. (25) into Eq. (22), and simplifying the results yields

$$\mathbf{P}_k^+ = \left(\mathbf{I} - \mathbf{K}_k \mathbf{H}_k\right) \mathbf{P}_k^- + \mathbf{S}_k^+ \mathbf{W} \boldsymbol{\gamma}_k^T \mathbf{K}_k^T \qquad (26)$$

The special DKF (SDKF) algorithm has an analytical gain matrix, which is different from the gain matrix in reference (Karlgaard & Shen, 2011). The SDKF algorithm includes Eqs. (16), (18), (23), (25), (17), (19) and (26) with initial conditions $\hat{\mathbf{x}}_0^+ = E[\mathbf{x}_0]$, $\mathbf{P}_0^+ = E[\mathbf{e}_0^+ \mathbf{e}_0^{+T}]$, $\mathbf{S}_0^+ = \mathbf{0}$.

In the sensitivity-weighting matrix $\mathbf{W}$ in Eq. (21), every element in the leading diagonal of $\mathbf{W}$ is a weight of one parameter respect to total states. The sensitivity-weighting matrix in Karlgaard and Shen (Karlgaard & Shen, 2011) gives a sensitivity weight of each parameter respect to each state. So, the SDKF is only a special case of the DKF proposed by Karlgaard and Shen (Karlgaard & Shen, 2011).

## 5. Consider Kalman filter based on sensitivity matrix

Here, some results about sensitivity matrices $\mathbf{S}_k$ in Eqs. (18) and (19) and the cross-covariance matrices $\mathbf{C}_k$ in Eqs. (8) and (12) are given firstly, and a novel recursive form of the CKF based on the sensitivity matrix (SMCKF) is derived.

Postmultiplying Eqs. (18) and (19) by $\mathbf{P}_{pp}$ gives

$$\mathbf{S}_k^- \mathbf{P}_{pp} = \mathbf{\Phi}_{k/k-1} \mathbf{S}_{k-1}^+ \mathbf{P}_{pp} + \mathbf{\Psi}_{k/k-1} \mathbf{P}_{pp} \qquad (27)$$

$$\mathbf{S}_k^+ \mathbf{P}_{pp} = (\mathbf{I} - \mathbf{K}_k \mathbf{H}_k) \mathbf{S}_k^- \mathbf{P}_{pp} - \mathbf{K}_k \mathbf{N}_k \mathbf{P}_{pp} \qquad (28)$$

Noting that $\mathbf{\Phi}_{k/k-1}$ and $\mathbf{I} - \mathbf{K}_k \mathbf{H}_k$ are nonsingular in the Kalman filter and $\mathbf{P}_{pp}$ is a positive definite matrix, and comparing Eqs. (27) and (28) with Eqs. (8) and (12) gives

$$\mathbf{C}_k^- = \mathbf{S}_k^- \mathbf{P}_{pp} \qquad (29)$$

$$\mathbf{C}_k^+ = \mathbf{S}_k^+ \mathbf{P}_{pp} \qquad (30)$$

The basic equations of the SMCKF are derived by substituting Eqs. (29) and (30) into the basic equations of the CKF in section 3. The estimated state propagation equation of the SMCKF is Eq. (6), and the corresponding estimated state error propagation equation, by rewriting Eq. (7), is

$$\mathbf{P}_k^- = \mathbf{\Phi}_{k/k-1} \mathbf{\Gamma}_{k-1}^+ \mathbf{\Phi}_{k/k-1}^T + \mathbf{S}_k^- \mathbf{P}_{pp} \mathbf{S}_k^{-T} + \mathbf{G}_{k-1} \mathbf{Q}_{k-1} \mathbf{G}_{k-1}^T \qquad (31)$$

where $\mathbf{\Gamma}_{k-1}^+$ is defined as

$$\mathbf{\Gamma}_{k-1}^+ \triangleq \mathbf{P}_{k-1}^+ - \mathbf{S}_{k-1}^+ \mathbf{P}_{pp} \mathbf{S}_{k-1}^{+T} \qquad (32)$$

The gain matrix $\mathbf{K}_k$ in Eq. (9) is rewritten as

$$\mathbf{K}_k = (\mathbf{\Gamma}_k^- \mathbf{H}_k^T + \mathbf{S}_k^- \mathbf{P}_{pp} \mathbf{\gamma}_k^T)(\mathbf{H}_k \mathbf{\Gamma}_k^- \mathbf{H}_k^T + \mathbf{\gamma}_k \mathbf{P}_{pp} \mathbf{\gamma}_k^T + \mathbf{R}_k)^{-1} \quad (33)$$

The a posteriori state estimate updated by the measurement is Eq. (10), and the a posteriori state estimation error covariance in the SMCKF is

$$\mathbf{P}_k^+ = (\mathbf{I} - \mathbf{K}_k \mathbf{H}_k)\mathbf{\Gamma}_k^- + \mathbf{S}_k^+ \mathbf{P}_{pp} \mathbf{S}_k^{+T} + \mathbf{S}_k^+ \mathbf{P}_{pp} \mathbf{\gamma}_k^T \mathbf{K}_k^T \quad (34)$$

where $\mathbf{\Gamma}_k^-$ is defined as

$$\mathbf{\Gamma}_k^- \triangleq \mathbf{P}_k^- - \mathbf{S}_k^- \mathbf{P}_{pp} \mathbf{S}_k^{-T} \quad (35)$$

From the definitions of the new covariances $\mathbf{\Gamma}_{k-1}^+$ and $\mathbf{\Gamma}_k^-$ in Eqs. (32) and (35), we can obtain their corresponding propagation equations of $\mathbf{\Gamma}_k^-$ and $\mathbf{\Gamma}_k^+$ as follows

$$\mathbf{\Gamma}_k^- = \mathbf{\Phi}_{k/k-1} \mathbf{\Gamma}_{k-1}^+ \mathbf{\Phi}_{k/k-1}^T + \mathbf{G}_{k-1} \mathbf{Q}_{k-1} \mathbf{G}_{k-1}^T \quad (36)$$

$$\mathbf{\Gamma}_k^+ = (\mathbf{I} - \mathbf{K}_k \mathbf{H}_k)\mathbf{\Gamma}_k^- + \mathbf{S}_k^+ \mathbf{P}_{pp} \mathbf{\gamma}_k^T \mathbf{K}_k^T \quad (37)$$

It means that the SMCKF algorithm could work following equations (6), (18), (36), (33), (10), (19) and (37) with initial conditions $\hat{\mathbf{x}}_0^+ = E[\mathbf{x}_0]$, $\mathbf{P}_0^+ = E[\mathbf{e}_0^+ \mathbf{e}_0^{+T}]$, $\mathbf{S}_0^+ = \mathbf{0}$.

## 6. Equivalence between SMCKF and SDKF

In this section, it is demonstrated that the SMCKF and the SDKF are equivalent under the dynamic system (1) and (2) when the sensitivity-weighting matrix $\mathbf{W}$ in Eq. (25) is substituted by the covariance $\mathbf{P}_{pp}$ of the uncertain parameters.

Comparing the SMCKF and the SDKF, it is observed that they are the same in form, except that the new covariance $\mathbf{\Gamma}_k$ in SMCKF is different from the covariance $\mathbf{P}_k$ in SDKF. In fact, they have the same estimation when $\mathbf{W}$ is equivalent to $\mathbf{P}_{pp}$.

Assuming that the initial conditions are $\hat{\mathbf{x}}_0$, $\mathbf{P}_0$, $\mathbf{S}_0$, $\mathbf{P}_{pp}$, respectively. Note that $\hat{\mathbf{x}}_0^+ = \hat{\mathbf{x}}_0$, the uncertain parameters do not affect the initial value of $\hat{\mathbf{x}}_0^+$, hence the initial sensitivity is always $\mathbf{S}_0^+ = \mathbf{S}_0 = \mathbf{0}$, which is consistent with the cross-covariance matrices $\mathbf{C}_0^+ = 0$ in the CKF(Schmidt, 1966; Jazwinski, 1970; Tapley, et al., 2004). So, the corresponding initial covariance $\mathbf{\Gamma}_0$ of the SMCKF satisfies

$$\mathbf{\Gamma}_0 \triangleq \mathbf{P}_0 - \mathbf{S}_0 \mathbf{P}_{pp} \mathbf{S}_0^T = \mathbf{P}_0 \quad (38)$$

It implies that the two methods have the same algorithm formula when they have the same initial conditions. From section 4, it is known that the SMCKF, or the CKF, is a subset of the proposed DKF by Karlgaard and Shen (Karlgaard & Shen, 2011).

The sensitivity-weighting matrix in SDKF is known as a time-variant matrix respect to the uncertain parameters (Karlgaard & Shen, 2011), but how to select and obtain the sensitivity-weighting matrix is not discussed. The equivalence of the SMCKF and the SDKF means that the a priori covariance of the uncertain parameters could be selected as the sensitivity-weighting matrix in the SDKF.